\documentclass[12pt]{article}
\usepackage{amsmath}
\usepackage{graphicx}
\input epsf
\begin{document}
\thispagestyle{empty}
\begin{center}

{\Large\bf Positivity constraints for lepton polarization \vskip 0.4cm

in neutrino deep inelastic scattering}\\

\vskip1.4cm
Claude Bourrely and Jacques Soffer  
\vskip 0.3cm
Centre de Physique Th\'eorique\footnote{UMR 6207 - Unité Mixte de Recherche 
du CNRS et des Universités Aix-Marseille I, Aix-Marseille II et de
l'Université du Sud Toulon-Var - Laboratoire affilié à la FRUMAM}, 
CNRS-Luminy, \\
Case 907, F-13288 Marseille Cedex 9 - France \\ 
\vskip 0.5cm
Oleg V. Teryaev
\vskip 0.3cm
Bogoliubov Laboratory of Theoretical Physics, \\
Joint Institute for Nuclear Research, Dubna, 141980, Russia
\vskip 2cm
{\bf Abstract}\end{center}
We consider the spin polarization of leptons produced in neutrino and 
antineutrino nucleon deep inelastic scattering, via charged currents, 
and we study the positivity constraints on the spin components in a model
independent way. These results are very important, in particular in the
case of $\tau^{\pm}$ leptons, because the polarization information is crucial
in all future neutrino oscillation experiments.

\vskip 1cm
\noindent CPT-2004/P.001

\newpage

\section{Introduction}

Recent studies from neutrino oscillation experiments \cite{KSH,DC,SKK} provide
evidence for non-zero neutrino masses. Results from the Super-Kamiokande 
underground experiment \cite{SKK} measuring the atmospheric neutrino flux, 
suggest that muon neutrinos oscillate into tau neutrinos with
nearly maximal mixing. This $\nu_{\mu} \to \nu_{\tau}$ oscillation hypothesis
can be tested by means of $\tau$ production via $\nu_{\tau}$ scattering 
through charged current interactions, namely
\begin{equation}
\nu_{\tau}(\bar {\nu}_{\tau}) + N \to \tau^- (\tau^+) + X~~,
\label{dis}
\end{equation}
where $N$ is a nucleon target. This process will be studied with underground 
neutrino telescopes, such as AMANDA, ANTARES, NESTOR and BAIKAL \cite{NT}, 
as well as long-baseline neutrino oscillation experiments, such as ICARUS, 
MINOS, MONOLITH and OPERA \cite{LBL}. Recently several authors have calculated
the $\tau$ production cross section for nuclear targets \cite{PY,KR}, but the
$\tau$ polarization should be also studied in order to estimate more
precisely the background events. This was the motivation for recent 
calculations of the $\tau$ polarization, which have been  
achieved in the framework of some particular models \cite{hag03,kln}, for deep 
inelastic scattering (\ref{dis}), but also for quasi-elastic scattering and 
resonance production. 

The relevance of positivity in spin physics, which puts strong restrictions on
spin observables in many areas of particle physics, has been already
emphasized \cite{JS} and the above process is one more example.
In this paper we show that the use of model independent positivity constraints 
reduces considerably the allowed region for the $\tau$ polarization. In the
next section we recall the kinematics, the general formalism
for deep inelastic scattering and the expressions for the components
of the $\tau$ polarization.
In Section 3, we exhibit the positivity conditions and our numerical results,
which have a direct relevance to the experiments mentioned above. 
Concluding remarks are given in Section 4 and some technical considerations
about the positivity of the hadronic tensor are given in the Appendix.

\newpage

\section{General formalism and kinematics}

In lepton nucleon deep inelastic scattering all the observables involve
the hadronic tensor of the nucleon $W_{\mu\nu}(p,q)$, where $p$, $k$ and
$k^{'}$ are the four momenta of the nucleon, incoming
$\nu_{\tau}$($\bar {\nu}_{\tau}$) and produced $\tau^-$ ($\tau^+$),
respectively, and $q = k - k^{'}$ is the momentum transfer. 
Since we consider the scattering of an unpolarized nucleon, using
Lorentz invariance and time reversal invariance, we can express 
$W_{\mu\nu}(p,q)$ in terms of five real structure functions
$W_i$ as follows \cite{LS,AJ,XJ},
\begin{eqnarray}
W_{\mu\nu}(p,q)&=&-g_{\mu\nu}W_{1}(\nu,q^{2})
+\frac{p_{\mu}p_{\nu}}{M^{2}}\,W_{2}(\nu,q^{2})
-i\epsilon_{\mu\nu\alpha\beta}\frac{p^{\alpha}q^{\beta}}
{2M^{2}}\,W_{3}(\nu,q^{2})\nonumber\\
&&+\frac{q_{\mu}q_{\nu}}{M^{2}}\,W_{4}(\nu,q^{2})
+\frac{p_{\mu}q_{\nu}+q_{\mu}p_{\nu}}{2M^{2}}\,W_{5}(\nu,q^{2})~.
\label{htensor}
\end{eqnarray}
Here $\epsilon_{\mu\nu\alpha\beta}$ is the total antisymmetric tensor with
$\epsilon_{0123} = +1$ and $W_3$ appears because of parity violation of weak
interactions. All structure functions, which are made dimensionless by
including appropriate mass factors, depend on two Lorentz scalars
$\nu = p\cdot q/M$ and $q^2 = -Q^2$ ($Q^2 > 0$), where $M$ is the
nucleon mass. In the laboratory frame, let us denote by
$E_{\nu}$, $E_{\tau}$ and $p_{\tau}$ the neutrino energy, $\tau$
energy and momentum, respectively and $\theta$ the scattering angle. We then
have $\nu = E_{\nu} - E_{\tau}$ and 
$Q^2=2E_{\nu}[E_{\tau}-p_{\tau}\cos{\theta}]-m^2_{\tau}$,
where $m_{\tau} = 1.777 \mbox{GeV}$ is the $\tau$ mass. Finally, the Bjorken 
variable $x$ is defined as $x= Q^2/2p\cdot q$ and the physical region is
$x_{min}\leq x \leq 1$, where $x_{min} = m_{\tau}^2/2 M (E_{\nu} - m_{\tau})$.
The unpolarized cross sections for deep inelastic scattering (\ref{dis}),
are expressed as
\begin{equation}
\label{cs}
\frac{d\sigma^{\pm}}{dE_{\tau}d\cos{\theta}}=\frac{G^2_F}{2\pi}
\frac{M_W^{4}p_{\tau}}{(Q^2+M_W^{2})^2}R_{\pm}~,
\end{equation}
where $G_F$ is the Fermi constant and $M_W$ is the $W$-boson mass. Here
\begin{eqnarray}
R_{\pm}&=&
\frac{1}{M}\,\bigg\{
\Big(2W_{1}+\frac{m_{\tau}^{2}}{M^{2}}\,W_{4}\Big)
\left(E_{\tau}-p_{\tau}\cos\theta\right)
+W_{2}\left(E_{\tau}+p_{\tau}\cos\theta\right)
\nonumber\\
&& \pm\frac{W_{3}}{M}\,\Big(E_{\nu}E_{\tau}+p_{\tau}^{2}
-(E_{\nu}+E_{\tau})p_{\tau}\cos\theta\Big)
-\frac{m_{\tau}^{2}}{M}\,W_{5}\bigg\}~,
\label{cross}
\end{eqnarray}
where the $\pm$ signs correspond to $\tau^{\mp}$ productions.

Because of time reversal invariance, the polarization vector
$\overrightarrow P$ of the $\tau$ in its rest frame, lies in the
scattering plane defined by the momenta of the incident neutrino and the
produced $\tau$. It has a component $P_L$ along the direction of
$\overrightarrow {p_{\tau}}$ and a component $P_P$ perpendicular to 
$\overrightarrow {p_{\tau}}$, whose expressions are, in the laboratory
frame, \cite{hag03,kln,AJ}
\begin{eqnarray}
P_P\!\!\!&=&\!\!\! \mp\,\frac{m_{\tau}\sin\theta}{MR_{\pm}}
\bigg(2W_{1}-W_{2}\pm\frac{E_{\nu}}{M}\,W_{3}
-\frac{m_{\tau}^{2}}{M^{2}}\,W_{4}+\frac{E_{\tau}}{M}\,W_{5}\bigg)
,\\
P_L \!\!\!&=&\!\!\! \mp\,\frac{1}{MR_{\pm}}\bigg\{
\Big(2W_{1}-\frac{m_{\tau}^{2}}{M^{2}}\,W_{4}\Big)
\left(p_{\tau}-E_{\tau}\cos\theta\right)
+W_{2}\left(p_{\tau}+E_{\tau}\cos\theta\right)
\nonumber\\&&
\pm\frac{W_{3}}{M}\,\Big((E_{\nu}+E_{\tau})p_{\tau}
-(E_{\nu}E_{\tau}+p_{\tau}^{2})\cos\theta\Big)
-\frac{m_{\tau}^{2}}{M}\,W_{5}\cos\theta\bigg\}.
\label{polvec}
\end{eqnarray}
In addition, it is convenient to introduce also the degree of polarization 
defined as $P = \sqrt{P_P^2+P_L^2}$.
As previously the $\pm$ signs correspond to $\tau^{\mp}$ productions and
it is clear that if $W_3=0$, one has $R_{+}=R_{-}$ and $\tau^+$ and $\tau^-$
have opposite polarizations.
We also note that if one can neglect the mass of the produced lepton 
($m_{\tau}=0$), $P_P=0$, so such a lepton is purely left-handed, if negatively
charged, or purely right-handed, if positive.

\section{Positivity constraints and numerical results}

From Eq.~(\ref{htensor}) clearly the hadronic tensor $W_{\mu\nu}(p,q)$ is
Hermitian 
\begin{equation}
W_{\mu\nu}(p,q) = W_{\nu\mu}^*(p,q)~,
\label{her}
\end{equation}
and semi-positive. This last property implies that
\begin{equation}
a^*_{\mu}W_{\mu\nu}(p,q)a_{\nu} \geq 0 ~,
\label{gpos}
\end{equation}
for {\it any} complex 4-vector $a_{\mu}$. The 4x4 matrix representation of 
$W_{\mu\nu}(p,q)$ in the laboratory frame where $p=(M,0,0,0)$
and $q=(\nu,\sqrt{\nu^2+Q^2},0,0)$ reads
$\left ( \begin {array}{c c} M_{1} & 0 \\  0 & M_{0} \end {array} \right)$
 where $M_1$ and $M_0$ are the following 2x2 Hermitian matrices
\begin {eqnarray} 
M_1 = \left ( \begin {array}{cc} -W_1+W_2+\frac{\nu^2}{M^2}W_4+
\frac{\nu}{M}W_5 & 
\frac{\sqrt{\nu^2+Q^2}}{M}(\frac{\nu}{M}W_4+\frac{1}{2}W_5) \\ 
\frac{\sqrt{\nu^2+Q^2}}{M}(\frac{\nu}{M}W_4+\frac{1}{2}W_5) & 
W_1+\frac{\nu^2+Q^2}{M^2}W_4 \end {array} \right)~,
\label{M10}
\end {eqnarray}
and
\begin {eqnarray}
M_0 = \left ( \begin {array}{cc} W_1 & \frac{-i\sqrt{\nu^2+Q^2}}{2M}W_3 \\ 
\frac{+i\sqrt{\nu^2+Q^2}}{2M}W_3  & W_1 \end {array} \right)~.
\label{M13}
\end {eqnarray}
The {\it necessary and sufficient conditions} for $W_{\mu\nu}(p,q)$ to satisfy
inequality~(\ref{gpos}) are that all the principal minors of $M_1$ and $M_0$
should be positive definite. So for the diagonal elements we have three
inequalities linear in the $W_i$'s namely
\begin{equation}
W_1 \geq 0 ~,
\label{W1}
\end{equation}
\begin{equation}
-W_1+W_2+\frac{\nu^2}{M^2}W_4+\frac{\nu}{M}W_5 \geq 0 ~,
\label{M11}
\end{equation}
\begin{equation}
W_1+\frac{\nu^2+Q^2}{M^2}W_4 \geq 0 ~,
\label{M22}
\end{equation}
and from the 2x2 determinants of $M_0$ and $M_1$ we get two inequalities
quadratic in the $W_i$'s namely
\begin{equation}
W_1^2 \geq \frac{\nu^2+Q^2}{4M^2}W_3^2 ~,
\label{W132}
\end{equation}
or equivalently
\begin{equation}
W_1 \geq \frac{\sqrt{\nu^2+Q^2}}{2M}|W_3| ~,
\label{W13}
\end{equation}
and
\begin{eqnarray}
\left(-W_1+W_2+\frac{\nu^2}{M^2}W_4+\frac{\nu}{M}W_5 \right) \left(
W_1+\frac{\nu^2+Q^2}{M^2}W_4 \right)
\nonumber \\ 
\geq \frac{\nu^2 + Q^2}{M^2} \left( \frac{\nu}{M}W_4+\frac{1}{2}W_5 \right)^2~.
\label{quad}
\end{eqnarray}
By imposing the last condition, only one of the two inequalities (\ref{M11})
 or (\ref{M22}) is needed, the other one follows automatically.
Since the hadronic tensor $W_{\mu\nu}(p,q)$ allows the construction of the 
scattering amplitudes for a vector-boson nucleon Compton scattering process,
the five structure functions $W_i$ are related to the five s-channel helicity 
amplitudes, which survive in the forward direction. As a special case
in Eq.~(\ref{gpos}), if one takes for $a_{\mu}$ the polarization vectors of
the vector-boson, the nucleon being unpolarized, these amplitudes are
\begin{equation}
M(h',h) = \epsilon_{\mu}^{*}(h') W_{\mu\nu} \epsilon_{\nu}(h)~,
\label{amp}
\end{equation} 
where $h$ and $h'$ are the helicities of the initial and final 
vector-boson, respectively \footnote{For a complete study of deep inelastic
scattering with a polarized nucleon, in terms of fourteen structure functions,
see Ref.~\cite{XJ}.}. 
The positivity conditions reflect the fact that the forward amplitudes, which
are indeed cross sections, must be positive. The linear conditions
correspond to the polarized vector-boson scattering, with longitudinal,
transverse or scalar polarizations and the quadratic condition~(\ref{quad}),
is a Cauchy-Schwarz inequality which corresponds to the scalar-longitudinal
interference.
The above set of positivity constraints might appear to be different from the
ones derived earlier \cite{LY,DdR}, but this is not the case as we will
discuss in the Appendix.

In order to test the usefulness of these constraints to restrict the
allowed domains for $P_P$ and $P_L$, we proceed by the following method,
without refering to a specific model for the $W_i$'s. We generate randomly the
values of the $W_i$'s, in the ranges [0,+1] for $W_1$ and $W_2$, which are
clearly positive and [-1,+1] for $i=3,4,5$. The most trivial
positivity constraints are $R_{\pm} \geq 0$, but in fact they are too weak
and do not imply the obvious requirements $|P_L| \leq 1$ and $|P_P| \leq 1$
or $P \leq 1$ \footnote{ Note that in the trivial case where 
$W_3=W_4=W_5=0$, $R \geq 0$ implies $P \leq 1$.}. So we first impose
$R_{\pm} \geq 0$ and $P \leq 1$ for different values of $E_{\nu}$, $Q^2$ and
$x$ and as shown in Fig.~1, for $\tau^+$ production, the points which satisfy
these constraints are represented by grey dots inside the disk,
$P_{L}^2 + P_{P}^2 \leq 1$. 
If we now add the non trivial positivity constraints Eqs.(10-15), which also
garantee that $P\leq 1$, we get the black dots, giving a much smaller area.
In Fig.~1, the top row corresponds to $E_{\nu}= 10 \mbox{GeV}$ and
$Q^2 = 1\mbox{GeV}^2$, the row below to $E_{\nu}= 10 \mbox{GeV}$
and $Q^2 = 4\mbox{GeV}^2$ and the next two rows to $E_{\nu}= 20 \mbox{GeV}$ and
$Q^2 = 1,4 \mbox{GeV}^2$. Going from left to right $x$ increases from a value
close to its minimum to 0.9. It is interesting to note that the black allowed
area increases with $Q^2$ and becomes smaller for increasing incident energy
and increasing $x$. For $\tau^-$ production, the corresponding areas are
obtained by symmetry with respect to the center of the disk. 
For increasing $x$, since $P_L$ is more and more restricted to values close to
+1 for $\tau^+$ (-1 for $\tau^-$), it is striking to observe that the non
trivial positivity constraints lead to a situation where the
$\tau^+$ ( $\tau^-$) is almost purely right-handed (left-handed), although
it has a non zero mass.

Another way to present our results is seen in Fig.~2, which shows the upper and
lower bounds from the non trivial positivity constraints for a given incident
energy and different $x$ values, versus $Q^2$.
These bounds are obtained by selecting the larger and smaller allowed values
of $P_L$ and $P_P$, when the $W_i$'s are varied for a fixed bin
of $E_{\nu}$ and $x$. 
We also indicate the scattering angle which increases with $Q^2$ and we recall
that for $\theta=0$ we have $P_P=0$ (see Eq.~(4)).

Finally we have tested the effect of some approximate relations among the 
$W_i$'s, which have been proposed in the literature. 
First, as an example for a particular kinematic situation we show in Fig.~3 
the effect of imposing the Callan-Gross relation \cite{CG},
namely $Q^2W_1=\nu^2W_2$. It further reduces both the grey dots and the black
dots areas, since this has to be compared with the first row of Fig.~1.
For the same kinematic situation we also show in Fig.~4, the effect of the
Albright-Jarlskog relations \cite{AJ}, namely $MW_1=\nu W_5$ and $W_4 =0$,
and we observe again that the allowed regions are much smaller.
These examples illustrate the fact that a more precise knowledge of the
structure functions $W_i$'s, will certainly further restrict the domains
shown in Fig.1.

\section{Concluding remarks}

We have shown in this paper that the positivity conditions on the hadronic
tensor of the nucleon $W_{\mu\nu}(p,q)$, is essential to reduce
the allowed values for the
$\tau^{\pm}$ polarization in neutrino deep inelastic scattering. We have not
used a specific model and we have considered only a few kinematic situations,
which are relevant for the long baseline neutrino oscillation
experiments, but they can be easily applied to other kinematic ranges and in
the framework of any given model.
They are less usefull for ultra high neutrino energies, because in this case
$\theta \simeq 0$, so $P_P \simeq 0$ and $P_L \simeq \pm 1$ for $\tau^{\mp}$.
The universality of $W_{\mu\nu}(p,q)$, which occurs in processes we have not
studied here (i.e. quasi-elastic scattering etc...), also increases the
importance of these positivity constraints.

\section{Appendix}

The positivity conditions on $W_{\mu\nu}(p,q)$ were first obtained in
Refs.~\cite{LY,DdR} and they were reported in Refs.~ \cite{LS,AJ} under
a slightly different form due to the use of our definition of 
$W_{\mu\nu}(p,q)$, which differs from that of Ref.~\cite{DdR}.
Moreover in Ref.~\cite{DdR} instead of the laboratory system, they were using
a frame where $q$ is purely space-like.
Although from covariance one expects the equivalence of the different sets
of conditions, it seems natural to show it explicitely. Let us
consider the frame where $p=(M\sqrt{1+\nu^2/Q^2},-\nu M/\sqrt{Q^2},0,0)$ and
$q=(0,\sqrt{Q^2},0,0)$.
The 4x4 matrix representation of $W_{\mu\nu}(p,q)$ is very similar to the case
of the laboratory frame, since it reads 
$\left ( \begin {array}{c c} M_{2} & 0 \\  0 & M_{0} \end {array} \right)$
where $M_2$ is
\begin {eqnarray} 
M_2 = \left ( \begin {array}{c c} -W_1+ (1 + \frac{\nu^2}{Q^2} )W_2 & 
\frac{\sqrt{\nu^2+Q^2}}{2M}(W_5-\frac{2M\nu}{Q^2}W_2) \\ 
\frac{\sqrt{\nu^2+Q^2}}{2M}(W_5-\frac{2M\nu}{Q^2}W_2)& 
W_1+\frac{\nu^2}{Q^2}W_2+\frac{Q^2}{M^2}W_4-\frac{\nu}{M}W_5\end {array}
\right)~,
\label{M20}
\end {eqnarray}
and $M_0$ was given in (\ref{M13}).
The momenta $p$ and $q$ defined in the two reference frames are related by a
Lorentz transform, so the matrix elements of $M_1$ and $M_2$ are simply
related. Moreover one can check that, first, 
\begin{equation}
\mbox{det}(M_1)=\mbox{det}(M_2)~,
\label{end}
\end{equation}
second, the difference of the diagonal elements of $M_1$ and $M_2$ is the same
and these diagonal elements must be both either positive or negative,
due to Eq.~(\ref{end}). So in order to establish the equivalence of the
positivity conditions in the two reference frames, a simple calculation proves
that the two inequalities (\ref{M11}) or (\ref{M22}) imply
\begin{equation}
-W_1+ (1 + \frac{\nu^2}{Q^2} )W_2 \geq 0
\end{equation}
or
\begin{equation}
W_1+\frac{\nu^2}{Q^2}W_2+\frac{Q^2}{M^2}W_4-\frac{\nu}{M}W_5 \geq 0~.
\end{equation}
\vskip 0.3cm \noindent {\bf Acknowledgments:} 
We thank V. A. Naumov and K. Kuzmin for some discussions and correspondance.
O. V. Teryaev is grateful to the Centre de Physique Th\'eorique, where part
of this work was done, and CNRS for financial support. His work was partially 
supported by RFBR (Grant 03-02-16816).

\newpage

\begin{figure}[t]
  \vspace*{-21mm}
\hspace*{-0.18\textwidth}
\leavevmode {\epsfxsize=17.cm \epsfysize=24.cm  \epsffile{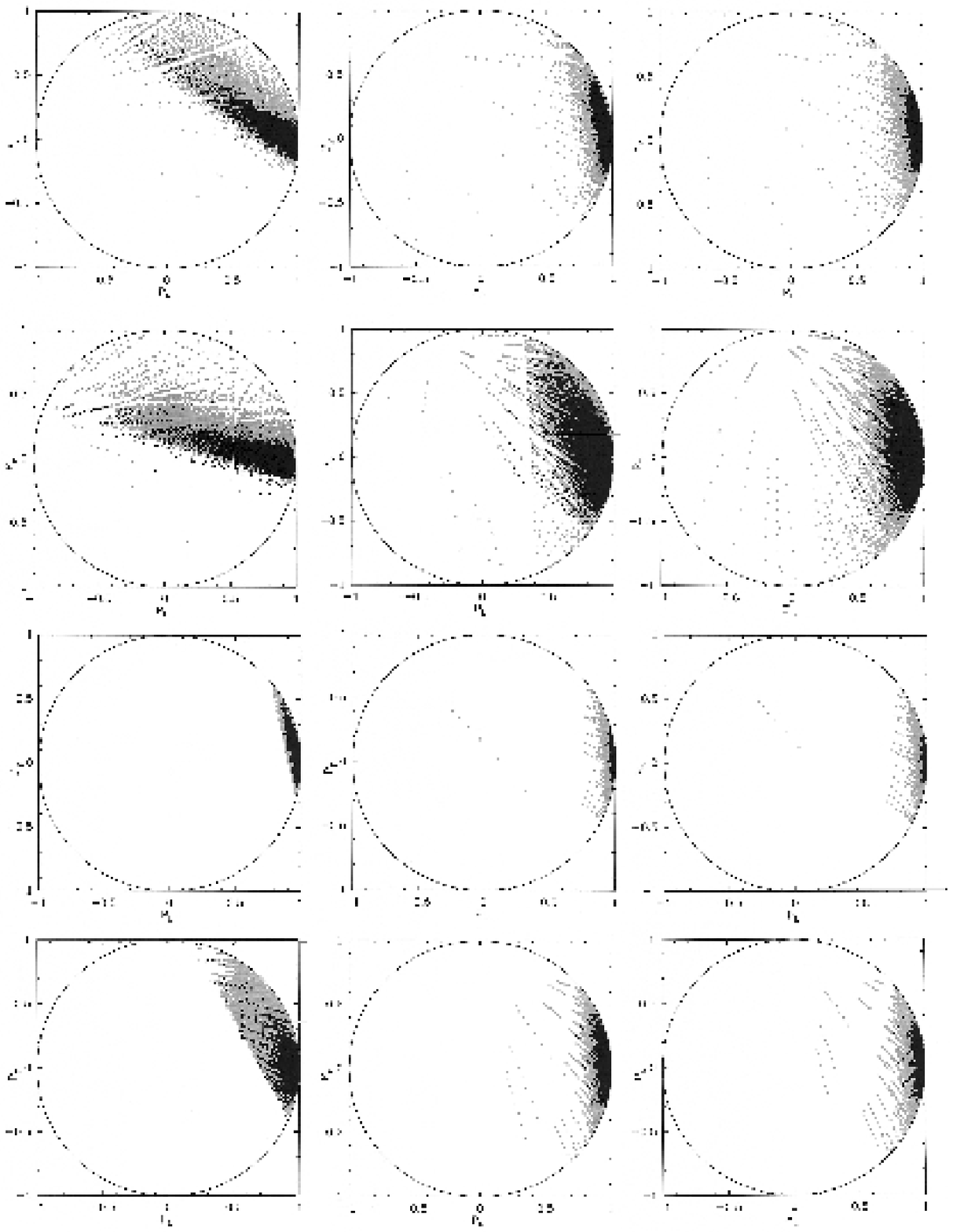}}
  \vspace*{-55mm}
\caption{\label{fig:1} For $\tau^+$ production, $P_P$ versus $P_L$ in a
domain limited by $R_{+} \geq 0$, $P \leq 1$ (grey area)
plus non trivial positivity constraints (black area).
From top to bottom and left to right,
$E_{\nu} = 10\mbox{GeV},~Q^2 = 1\mbox{GeV}^2,~x = 0.25, 0.6, 0.9$,
$E_{\nu} = 10\mbox{GeV},~Q^2 = 4\mbox{GeV}^2,~x = 0.4, 0.6, 0.9$,
$E_{\nu} = 20\mbox{GeV},~Q^2 = 1\mbox{GeV}^2,~x = 0.25, 0.6, 0.9$,
$E_{\nu} = 20\mbox{GeV},~Q^2 = 4\mbox{GeV}^2,~x = 0.25, 0.6, 0.9$.}
\end{figure}
\newpage

\begin{figure}[thb]
  \vspace*{-25mm}
\hspace*{-0.20\textwidth}
\leavevmode {\epsfxsize= 19.cm \epsfysize=26.cm \epsffile{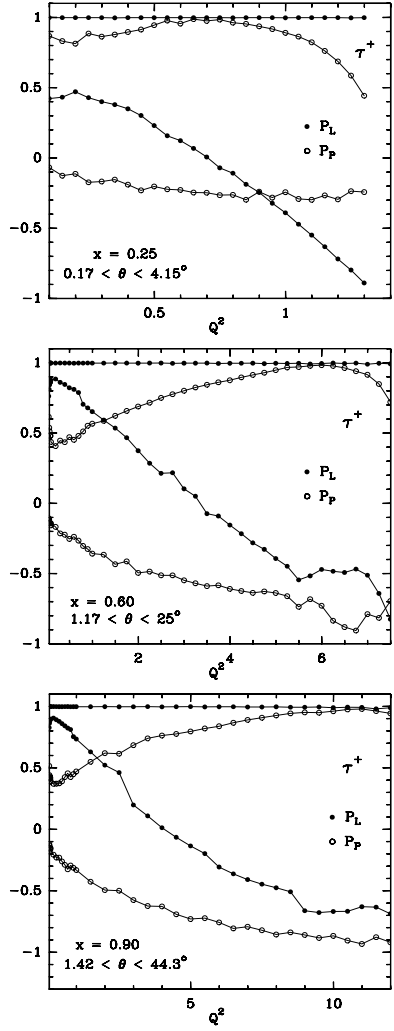}}
\vspace*{-70mm}
\caption{\label{fig:2}For $\tau^+$ production, upper and lower bounds 
on $P_P$ (open circles) and  $P_L$ (full circles) as a function of $Q^2$ for 
$E_{\nu} = 10\mbox{GeV}$ and $x = 0.25, 0.6, 0.9$.}
\end{figure}

\newpage
\begin{figure}[thb]
  \vspace*{-21mm}
\hspace*{-0.18\textwidth}
\leavevmode {\epsfxsize=17.cm \epsfysize=24.cm \epsffile{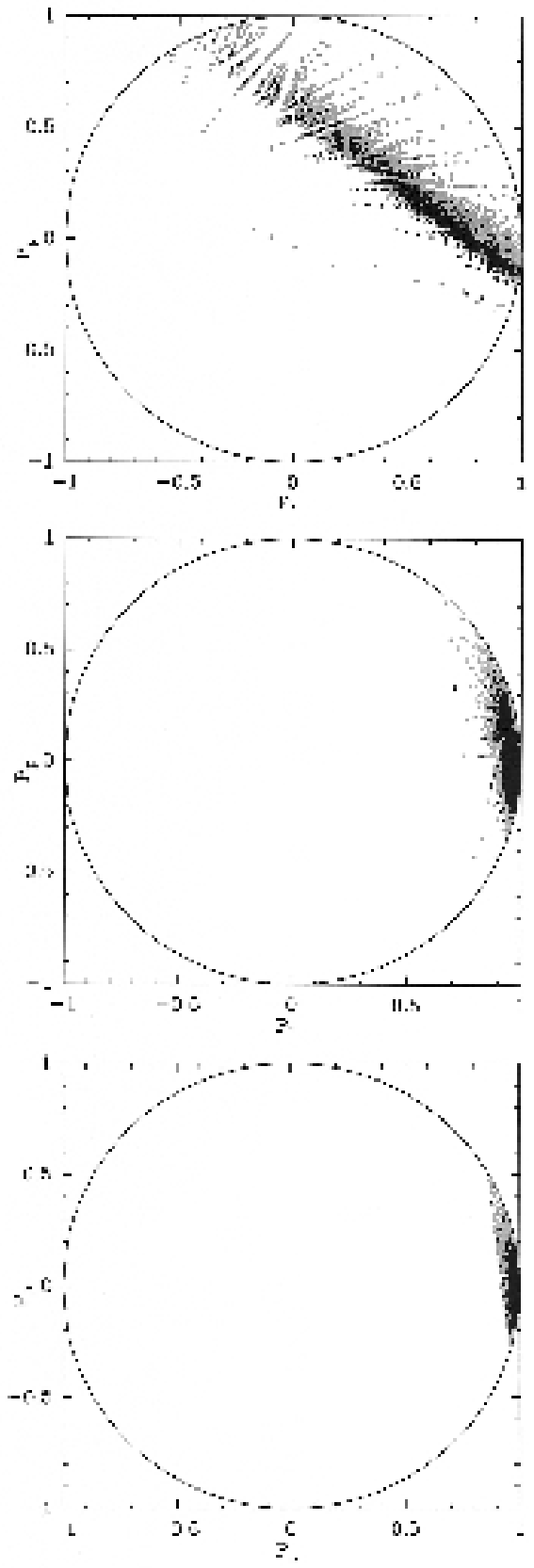}}
\vspace*{-55mm}
\caption{\label{fig:3}For $\tau^+$ production, $P_P$ versus $P_L$ in a domain 
limited by $R_{+} \geq 0$, $P \leq 1$ assuming the Callan-Gross relation
(grey area) plus non trivial positivity  constraints (black area).
$E_{\nu} = 10\mbox{GeV},~Q^2 = 1\mbox{GeV}^2$, from top to bottom,
$x = 0.25, 0.6, 0.9$.}
\end{figure}

\newpage
\begin{figure}[thb]
  \vspace*{-21mm}
\hspace*{-0.18\textwidth}
\leavevmode {\epsfxsize= 17.cm \epsfysize=24.cm \epsffile{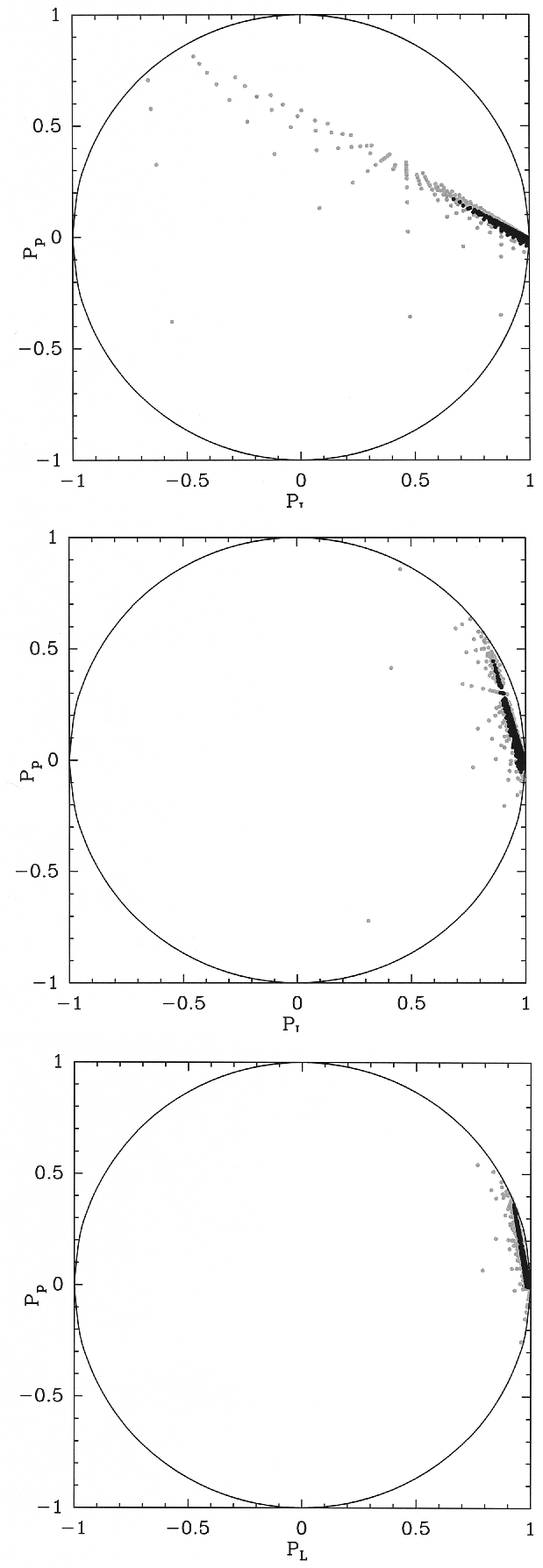}}
\vspace*{-55mm}
\caption{\label{fig:4}For $\tau^+$ production, $P_P$ versus $P_L$ in a domain 
limited by $R_{+} \geq 0$, $P \leq 1$ assuming the Albright-Jarlskog relations
(grey area) plus non trivial positivity 
constraints (black area).
$E_{\nu} = 10\mbox{GeV},~Q^2 = 1\mbox{GeV}^2$, from top to bottom,
$x = 0.25, 0.6, 0.9$.}
\end{figure}

\end{document}